# Data-driven ANN model for estimating unfrozen water content in the thermo-hydraulic simulation of frozen soils

Mingpeng Liu [1,*], Peizhi Zhuang [2], Raul Fuentes [3]

1. Ph. D. Candidate, Institute of Geomechanics and Underground Technology, RWTH Aachen University, Germany. Email: Liu@gut.rwth-aachen.de

2. Professor. School of Qilu Transportation, Shandong University, Jinan, 250061, China. Email: zhuangpeizhi@sdu.edu.cn

3. Professor. Institute of Geomechanics and Underground Technology, RWTH Aachen University, Germany. Email: raul.fuentes@gut.rwth-aachen.de

* **Corresponding author**:

Mingpeng Liu

Institute of Geomechanics and Underground Technology, RWTH Aachen University, Germany.

Email address: Liu@gut.rwth-aachen.de

# Data-driven ANN model for estimating unfrozen water content in the thermo-hydraulic simulation of frozen soils

## Abstract

This study integrates a data-driven model for estimating the unfrozen water content into the thermo-hydraulic coupling simulation of frozen soils. An artificial neural network (ANN) was employed to develop this data-driven model using a dataset from the literature. Thereafter, a numerical algorithm was developed to implement the data-driven model into the thermo-hydraulic simulation. In the numerical algorithm, the frozen and unfrozen zones are distinguished first according to the freezing temperature, where the unfrozen water at frozen nodes is updated using the ANN model. Subsequently, discretized hydraulic and thermal equations are solved sequentially and iteratively using Newton-Raphson method until the temperature and unfrozen water content satisfy the tolerance simultaneously. Horizontal and vertical freezing experiments are used to verify the reliability of the proposed algorithm. The computed variations in temperature, total water, unfrozen water, and ice content achieve good agreements with measured data. Some key features of frozen soils, such as water migration and ice formation, and the increase in total water content, are reproduced by the developed algorithm. Additionally, the comparison between the ANN model and existing empirical equations for determining unfrozen water content demonstrates that the ANN model offers a better performance.

## 1. Introduction

Frozen soils, including permafrost and seasonally frozen ground, are widely distributed in cold regions, covering approximately 24% of the Northern Hemisphere and 15% of the total land surface (Lai et al., 2014; Obu, 2021). Under continuous cooling of air temperature during the coldest months of the year, unfrozen liquid water becomes ice as depicted in Fig. 1 and the frozen front gradually penetrates deeper underground. With the onset of spring and rising temperatures, thaw settlement occurs in the active layer as ground ice melts, weakening soil shear strength. This freeze-thaw cycles in the active layer pose a significant damage threat to engineering infrastructure such as tunnels, bridges, pavements, and embankments (Thomas et al., 2009). The significant issue for subsurface soil suffering freeze-thaw action is to understand clearly the multi-faceted coupling processes, involving temperature evolution, moisture migration, and phase transition. These interactions in frozen soil are critical for infrastructure safety in cold regions.

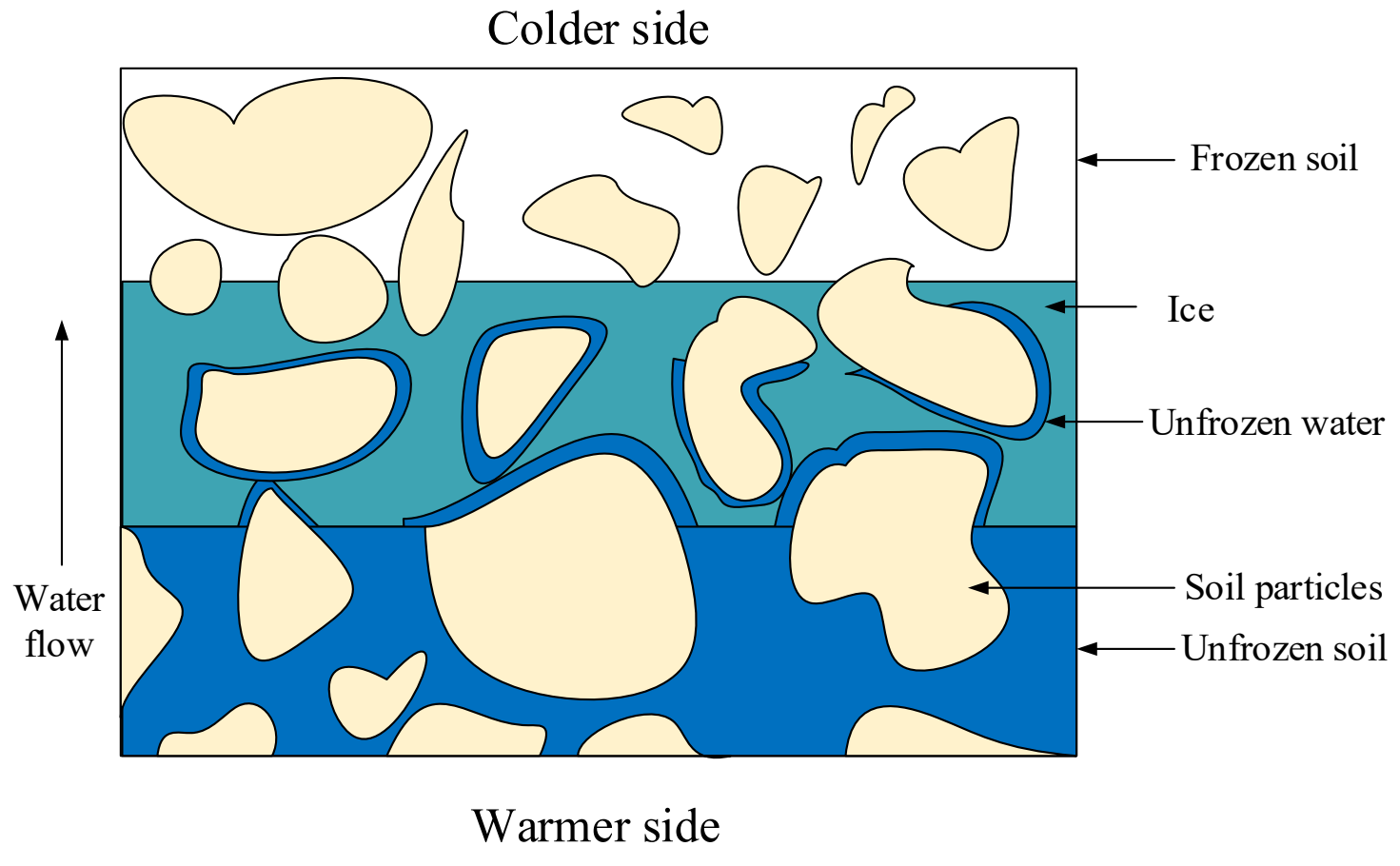

**Fig. 1**. Schema of the freezing soil.

Numerous laboratory experiments have been conducted to investigate the mechanisms of heat and water transfer in freezing soil. From these, it is clear that the temperature gradients drive the moisture movement in frozen soils (Hoekstra, 1966; Sweidan et al., 2022; Teng et al., 2019). Kemper (1960) proved the existence of a thin liquid water film at the surface of soil particles. This film coexists with ice water and acts as the transport media that allows water migration. Zhou et al. (2014) pointed out that this coexistence of water and ice significantly impacts the hydraulic, thermal, and mechanical properties of the soil. For example, ice alters soil hydrology by reducing hydraulic conductivity (Iwata et al., 2010). During the moisture migration, pore ice on the cold side continues to grow as incoming liquid water freezes, which may lead to an increase in total water content in the colder region (Hoekstra, 1966; Watanabe et al., 2011).

To simulate these complex thermo-hydraulic coupling process in frozen soils, numerical methods have been developed. Harlan (1973) introduced a numerical model for computing the variation of temperature and the redistribution of moisture during soil freezing. Afterwards, numerous mathematical models have been proposed (Liu and Sun, 2025; Lv et al., 2025; Suh and Sun, 2022) (Bekele et al., 2017; Sweidan et al., 2022; Vitel et al., 2016). Among these numerical models, a key aspect is the relationship between temperature and unfrozen water content, characterized by the soil freezing characteristic curve (SFCC) that serves for coupling the thermal transfer equation and hydraulic equation. Typically-used approaches for predicting the SFCC include power functions (Anderson and Tice, 1972; Osterkamp and Romanovsky, 1997), exponential functions (McKenzie et al., 2007; Michalowski, 1993; Stuurop et al., 2021), Clausius–Clapeyron equation (Kung and Steenhuis, 1986; Kurylyk and Watanabe, 2013; Loch, 1978), freezing point depression function (Cary and Mayland, 1972; Gray et al., 2001; Smirnova et al., 2000), to name but a few, and other physical-based models (Chen et al., 2022; Dall'Amico et al., 2011). These methods usually used specific mathematic expressions with predefined parameters, which however exhibit some limitations, e.g., the discrepancies with experimental observations and the necessity for calibrating parameters for each type of soil. These limitations have motivated scholars to develop other methods for estimating unfrozen water content.

With the advancement of machine learning techniques, studies have explored data-driven models to estimate it (Zheng et al., 2025; Zhou et al., 2024). Ren et al. (2023) and Li et al. (2024a) used artificial neural network (ANN) to predict experimental unfrozen water content using data from the literature. Nartowska and Sihag (2024) and Li et al. (2024b) compared the performance of random forest (RF), extreme gradient boosting (XGBoost), light gradient boosting machine (LightGBM), K-nearest neighbors (KNN), support vector regression (SVR), and ANNs to identify the most effective models. Their results demonstrate that the ensemble algorithms achieved the better performance. To achieve a higher accuracy, Park et al. (2025) proposed to estimate unfrozen water content using a pedotransfer function implemented with XGBoost. Their findings indicated that the data-driven approaches outperform traditional empirical equations and present the advantage of surrogating the experimental relationship. Despite these advances, current data-driven models merely concentrate on predicting unfrozen water but are not extended to the thermo-hydraulic process of frozen soils, which heavily hinders the practical utilization of these advanced prediction models. Research on integrating data-driven models into thermo-hydraulic simulation remains limited, and an effective coupling algorithm for seamless implementation is still lacking.

This study aims therefore at implementing a data-driven model to estimate unfrozen water content into the thermo-hydraulic coupling simulation of frozen soils. The structure of this paper is as follows: Section 2 introduces the development of the data-driven model for estimating unfrozen water content. Section 3 illustrates the developed numerical algorithm, including the formulation of thermal and hydraulic transfer equations along with their discretization forms, and the algorithm to incorporate the data-driven model. Section 4 verifies the proposed numerical algorithm with experimental cases in temperature, total water, unfrozen water, and ice content. Section 4 also compares the performance of the data-driven model with empirical equations to assess its accuracy and reliability. Finally, Section 5 draws conclusions of this work.

# 2. Data-driven model for estimating unfrozen water content

## 2.1 Data sources

Ren et al. (2023) identified four important factors when estimating the SFCC: specific surface area (SSA), dry density ($\rho_d$), initial volumetric water content ($\theta_0$), and obviously temperature ($T$). In their framework, SSA and $\rho_d$ are used to identify soil types and properties. They compiled a dataset consisting of 73 groups of experimental cases reported in literature, providing detailed descriptions of the four key factors and corresponding unfrozen water content (shown in Table 1 excluding the last row). Additionally, experimental data from Zhou et al. (2014) (the last row in Table 1) is also supplied in the dataset, which is subsequently used to validate the numerical algorithm conducted in this work. The table shows that a wide spectrum of soils, testing conditions and temperature ranges are covered.

**Table 1**. Experimental information for the database of unfrozen water content.

| References | Type of soils | Group of tests | Test conditions for SFCC | Temperature range (°C) |
|---|---|---|---|---|
| Smith and Tice (1988) | Clay, silt, loam, kaolinite, gravel, bentonite, hectorite, ash | 25 | Thawing | -15.11 ~ 0 |
| Suzuki (2004) | Light Clay | 1 | Thawing | -10 ~ -1.99 |
| Yoshikawa and Overduin (2005) | Silt and clay | 2 | Freezing | -55.54 ~ 0 |
| Watanabe and Wake (2009) | Sand, silt loam, loam | 4 | Thawing | -14.69 ~ -0.01 |
| Ma et al. (2015) | Silt and clay | 2 | Thawing | -35.71 ~ -0.10 |
| Kruse and Darrow (2017) | Montmorillonite, kaolinite, illite, illite-smectit, chlorite, copper river | 6 | Both | -20 ~ -0.5 |
| Wang et al. (2020) | Silty clay | 1 | Thawing | -19.48 ~ -0.11 |
| Zhou et al. (2020) | Silt | 1 | Thawing | -18.90 ~ -0.08 |
| Lovell Jr (1957) | Silty clay, clay, clayey silt | 3 | - | -24.35 ~ -0.80 |
| Akagawa et al. (2012) | Kaolin, clay, ash, mudstone | 4 | Both | -20.54~ -0.14 |
| Wen et al. (2012) | Qinghai-Tibet silty clay | 1 | - | -15.05 ~ -0.49 |
| Zhou et al. (2015) | Silty clay | 1 | Both | -15.25 ~ -0.01 |
| Mu (2017) | Xi'an loess | 1 | Both | -9.82 ~ 0.10 |
| Chai et al. (2018) | Silty clay | 1 | Thawing | -14.97 ~ -0.10 |
| Mao et al. (2018) | Barcelona clayey silt | 1 | Freezing | -14.02 ~ -3.15 |
| Kong et al. (2020) | Sand, bentonite, and their mixture | 5 | Freezing | -22.01 ~ -0.04 |
| Li et al. (2020) | Silty clay, fine sand, medium sand | 3 | Both | -17.05 ~ -0.50 |
| Ren and Vanapalli (2020) | Canadian clay, Indian head till | 5 | Both | -18.70 ~ 0 |
| Teng et al. (2020) | Silica sand, silt, red clay | 3 | Both | -20.16 ~ -0.06 |
| Wang et al. (2021) | Silty clay, sandy loam, sand | 3 | Thawing | -19.45 ~ -0.11 |
| Zhou et al. (2014) | Mixture of sand, clay, and silt | 2 | Freezing | -2.98 ~ 0 |

## 2.2 Artificial neural network (ANN)

Fig. 2 illustrates a typical ANN, consisting of an input layer, multiple hidden layers, and an output layer. Each layer contains a set of neurons that are interconnected through weights and biases. The input data are first presented through the input layer and then past through the hidden layers to eventually predict values in the output layer. This process can be mathematically described by considering a feedforward propagation (Liu et al., 2024a):

$$\boldsymbol{y} = F(\boldsymbol{\omega x} + \boldsymbol{b}) \tag{1}$$

where $\boldsymbol{x}$ represents the input variables, $\boldsymbol{y}$ is the output; and $\boldsymbol{\omega}$ and $\boldsymbol{b}$ are the weight and bias matrix, respectively. $F$ represents the activation function. For the data-driven model in this study, the inputs are SSA, $\rho_d$, $\theta_0$, and $T$, while the output is volumetric unfrozen water content $\theta_w$. Accordingly, the data-driven model can be expressed as:

$$\theta_w = \mathcal{N}\mathcal{N}(\mathrm{SSA}, \theta_0, \rho_\mathrm{d}, T) \tag{2}$$

where $\mathcal{N}\mathcal{N}$ represents a neural network.

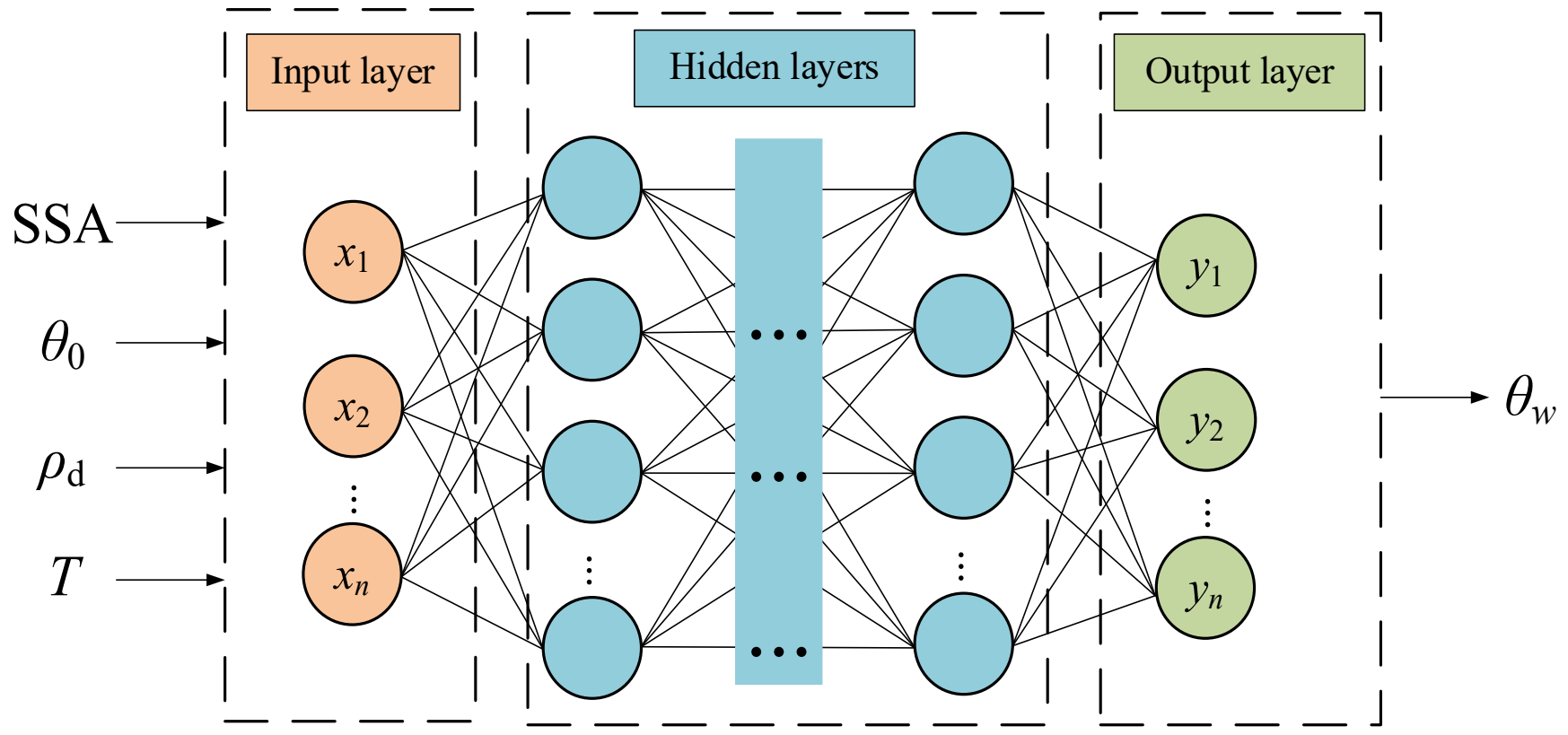

**Fig. 2**. The ANN model for estimating unfrozen water content.

Each group of experimental $T$-$\theta_w$ relationship contains several discretized points, resulting in a total of 1573 samples. The dataset is randomly divided into training and testing sets using a split ratio of 8 : 2. Data are normalized to the range of 0~1 using the maximal and minimal values within all experiments before training to minimize the impact of varying scales (Liu et al., 2024b):

$$x_\mathrm{norm} = \frac{x - x_\mathrm{min}}{x_\mathrm{max} - x_\mathrm{min}} \tag{3}$$

where $x$ is the raw input variables before normalisation, $x_\mathrm{norm}$ is the input variables after normalisation, $x_\mathrm{min}$, and $x_\mathrm{max}$ are the minimum and maximum values of the input variables, respectively.

Another critical aspect is the selection of optimal architecture and hyperparameters, including the number of hidden layers and neurons per layer. These hyperparameters were initially determined based on Bayesian optimization (Liu et al., 2024a), which aims to minimize the discrepancies between the training set and ANN predictions. Specifically, two hidden layers with specific neurons were set initially after the Bayesian optimization. Afterwords, we manually tuned the neurons in hidden layers and found by trial and error that 40 and 25 neurons for two hidden layers respectively achieved rge best model performance without overfitting in testing set.

To optimize the weight and bias matrices, we employed the Levenberg-Marquardt algorithm due to its high convergence rate towards optimal solutions, and the *Tanh* activation function is selected for the hidden layers. Additionally, to avoid overfitting, early stopping is implemented during the training process when the loss function value increases continuously in six consecutive epochs.

## 2.3 Performance of the data-driven model

The ANN model employs the mean square error (MSE) as the loss function. The coefficient of determination ($R^2$) is applied to evaluate the model performance:

$$R^2 = 1 - \frac{\sum_{i=1}^{n}(y_i - \tilde{y}_i)^2}{\sum_{i=1}^{n}(\tilde{y}_i - \bar{y})^2} \tag{4}$$

where $\tilde{y}_i$, $y_i$, and $\bar{y}$ denotes the true, predicted and mean true value of a neural network, respectively; $n$ denotes the number of samples.

The $R^2$ values for training and testing sets are 98.6% and 95.9%, respectively. Since Ren et al. (2023) quoted an $R^2$ value of 82% for the dataset, our prediction constitutes a significant improvement. This may be due to the fine-tuned network architecture and Levenberg-Marquardt algorithm. Fig. 3(a) shows the comparison between the predicted and the measured unfrozen water content in the testing set. The majority of samples are close to the 1:1 line, demonstrating that the ANN model accurately predicts the experimental unfrozen water content. Furthermore, Fig. 3(b) compares SFCCs predicted by the ANN model with those measured in experiments (Kong et al., 2020; Kruse and Darrow, 2017; Smith and Tice, 1988). The results indicate that the ANN model can satisfactorily capture the variation in unfrozen water with decreasing negative temperature for various types of soil and experimental conditions.

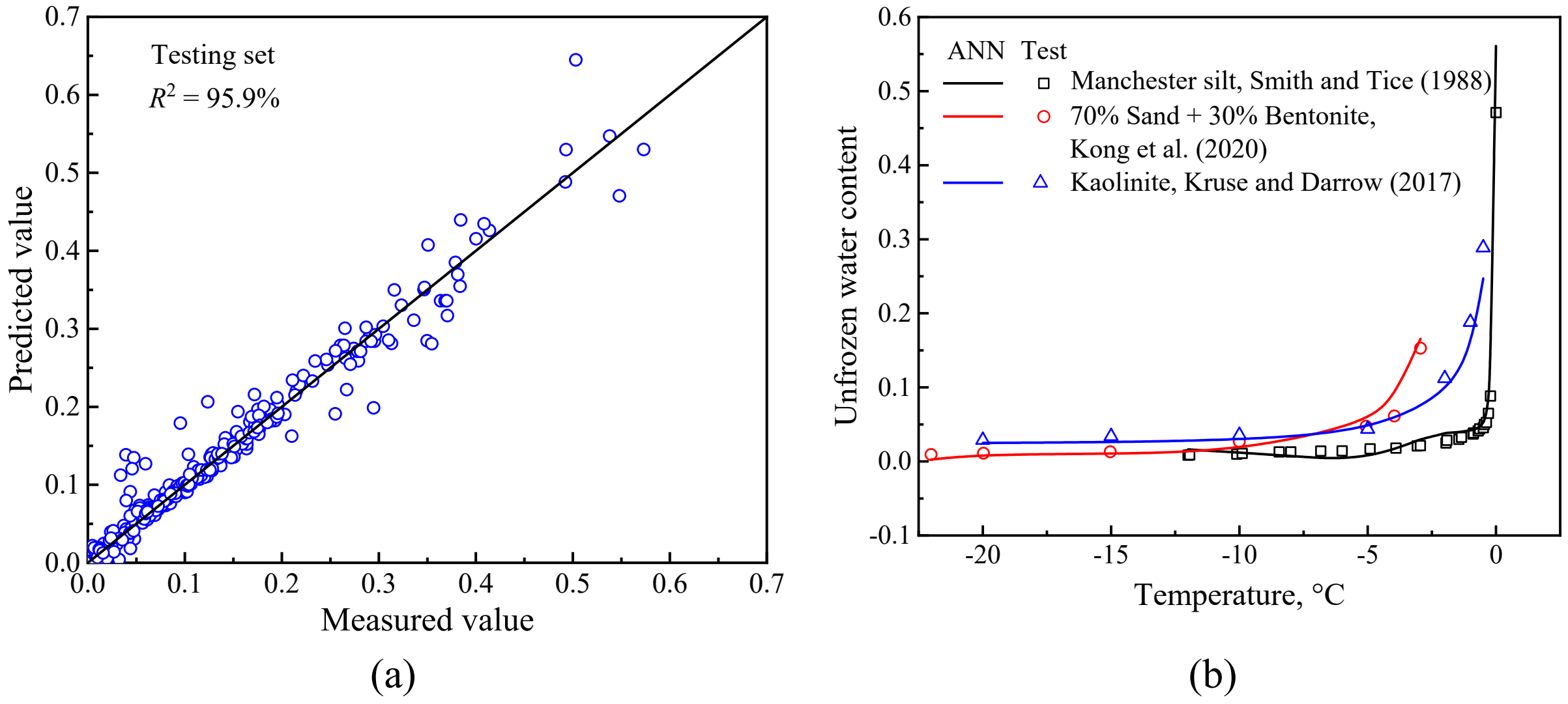

**Fig. 3**. Predicted versus measured (a) unfrozen water content in testing set, and (b) SFCCs in literature

# 3. Thermo-hydraulic simulation of frozen soils

The validated data-driven model is now integrated into the thermo-hydraulic simulation of frozen soils. For this purpose, the governing thermo-hydraulic equations are first introduced, followed by a detailed description of the numerical implementation.

## 3.1 Thermal transfer equation

Thermal transfer in unsaturated freezing soil can be expressed using Fourier's law. During the freezing process, the velocity of liquid water is sufficiently slow to ignore convective heat transfer (Fuchs et al., 1978). The thermal transfer equation can then be expressed as follows:

$$C\frac{\partial T}{\partial t} = \nabla \cdot (\lambda \nabla T) + L\rho_i \frac{\partial \theta_i}{\partial t} \tag{5}$$

where $C$ represents the volumetric heat capacity of soil; $\lambda$ denotes the thermal conductivity; $t$ is the time; $\rho_i$ is the density of ice; $\theta_i$ is the volumetric ice content; and $L$ is the latent heat of the phase change between water and ice. $\nabla$ is the Hamilton operator.

The volumetric heat capacity of soil $C$ can be expressed by:

$$C = \theta_w c_w \rho_w + \theta_i c_i \rho_i + (1 - \theta_s) c_s \rho_s \tag{6}$$

where $c_w$, $c_i$, $c_s$ represent the specific heat capacity of water, ice, and soil particles, respectively; $\rho_w$

and $\rho_s$ denote the density of water and soil particles, respectively; and $\theta_s$ is the water content at full saturation (determined from the void ratio).

For the thermal conductivity $\lambda$, Chen et al. (2022) pointed out that the traditional weighted arithmetic mean or weighted harmonic mean models exhibits significant deviations compared to experimental results. To address this limitation, this study proposes an improved weighted mean model:

$$\lambda = \beta[\theta_w\lambda_w + \theta_i\lambda_i + (1 - \theta_s)\lambda_s] \tag{7}$$

where $\lambda_w$, $\lambda_i$, and $\lambda_s$ represent the thermal conductivity of water, ice, and soil particles, respectively. $\beta$ is introduced as a thermal parameter to account for variations in different soil types. The effectiveness of this equation is further explained in Appendix B.

## 3.2 Hydraulic equation

We incorporate the resistance effect of pore ice on liquid water by utilizing Richards' equation. The water migration equation in frozen soil can be expressed as follows:

$$\frac{\partial\theta_w}{\partial t} + \frac{\rho_i}{\rho_w}\frac{\partial\theta_i}{\partial t} = \nabla\cdot[D(\nabla\theta_w + k)] \tag{8}$$

where $D$ and $k$ represent the diffusivity of water in the frozen soil and the permeability coefficient of water in unsaturated soil, respectively. The expressions of $D$ and $k$ can be obtained from the soil-water retention curve (SWRC). According to the VG water retention model (van Genuchten, 1980), $D$ and $k$ are defined as follows:

$$S = \frac{\theta_w - \theta_r}{\theta_s - \theta_r} \tag{9-1}$$

$$k = k_s S^{1/2}[1 - (1 - S^{1/m})^m]^2 \tag{9-2}$$

$$D = \frac{k}{c}\cdot I \tag{9-3}$$

$$c = \frac{\alpha m}{1-m}\left(1 - S^{\frac{1}{m}}\right)^m S^{\frac{1}{m}} \tag{9-4}$$

$$I = 10^{-10\theta_i} \tag{9-5}$$

where $S$ is the saturation degree; $\theta_r$ is the residual water content; $k_s$ is the saturated hydraulic conductivity; $\alpha$ and $m$ are the parameters in VG model; $I$ represents the impedance factor; $c$ is defined as the specific moisture capacity, typically calculated from the derivative of SWRC.

## 3.3 Spatial and temporal discretization

The highly nonlinear hydraulic and thermal transfer equations are typically solved using numerical discretization methods. In this study, the central difference is applied in the spatial domain, while the implicit backward difference is used in the temporal domain. The implicit backward difference significantly reduces time sensitivity. Taking 1-D heat transfer equation, the spatial and temporal domains are discretized into *N* nodes and *M* grids, respectively. Each item in Eq. (5) can be discretized as follows:

$$\left(\frac{\partial T}{\partial t}\right)_i = \frac{(T)_i^{n+1} - (T)_i^n}{\Delta t} \tag{10-1}$$

$$[\nabla \cdot (\lambda \nabla T)]_i = \left[\frac{\partial}{\partial x}\left(\lambda \frac{\partial T}{\partial x}\right)\right]_i = \frac{1}{\Delta x}\left[\lambda_{i+\frac{1}{2}}^{n+1} \frac{(T)_{i+1}^{n+1} - (T)_i^{n+1}}{\Delta x} - \lambda_{i-\frac{1}{2}}^{n+1} \frac{(T)_i^{n+1} - (T)_{i-1}^{n+1}}{\Delta x}\right] \tag{10-2}$$

$$\frac{\partial \theta_i}{\partial t} = \frac{(\theta_i)_i^{n+1} - (\theta_i)_i^n}{\Delta t} \tag{10-3}$$

where $\Delta t$ and $\Delta x$ denote the temporal and spatial interval, respectively. Superscript "$n$" represents the time step, and subscript "$i$" represents the spatial nodes. Subscript "$i + \frac{1}{2}$" and "$i - \frac{1}{2}$" in $\lambda$ obey $\lambda_{i+\frac{1}{2}}^{n+1} = \frac{1}{2}(\lambda_{i+1}^{n+1} + \lambda_i^{n+1})$ and $\lambda_{i-\frac{1}{2}}^{n+1} = \frac{1}{2}(\lambda_i^{n+1} + \lambda_{i-1}^{n+1})$, and applies similarly to other variables.

Accordingly, the implicit discretized form of Eq. (5) is written as:

$$a_i(T)_{i-1}^{n+1} + b_i(T)_i^{n+1} + d_i(T)_{i+1}^{n+1} = f_i \tag{11-1}$$

$$a_i = -\frac{\Delta t}{\Delta x^2}\frac{\lambda_{i-\frac{1}{2}}^{n+1}}{C_{i-\frac{1}{2}}^{n+1}}, b_i = 1 + \frac{\Delta t}{\Delta x^2}\left(\frac{\lambda_{i-\frac{1}{2}}^{n+1}}{C_{i-\frac{1}{2}}^{n+1}} + \frac{\lambda_{i+\frac{1}{2}}^{n+1}}{C_{i+\frac{1}{2}}^{n+1}}\right), d_i = -\frac{\Delta t}{\Delta x^2}\frac{\lambda_{i+\frac{1}{2}}^{n+1}}{C_{i+\frac{1}{2}}^{n+1}} \tag{11-2}$$

$$f_i = (T)_i^n + L\rho_i[(\theta_i)_i^{n+1} - (\theta_i)_i^n] / \left(\frac{C_i^{n+1} + C_i^n}{2}\right) \tag{11-3}$$

Eq. (11-1) is then assembled into a matrix form among all internal nodes:

$$\boldsymbol{J}_T : \boldsymbol{T} = \boldsymbol{F}_T \tag{12}$$

where $\boldsymbol{J}_T$ represents the Jacobian matrix for temperature, $\boldsymbol{T}$ is the matrix of nodal temperatures at the current time, and $\boldsymbol{F}_T$ is the residual matrix for temperature. Detailed expressions of each item in Eq. (12) are provided in the Appendix A.

Similarly, the discretized form of hydraulic equation is expressed as follows:

$$o_i(\theta_w)_{i-1}^{n+1} + p_i(\theta_w)_i^{n+1} + q_i(\theta_w)_{i+1}^{n+1} = r_i \tag{13-1}$$

$$o_i = -\frac{\Delta t}{\Delta x^2} D_{i-\frac{1}{2}}^{n+1}, p_i = 1 + \frac{\Delta t}{\Delta x^2}\left(D_{i+\frac{1}{2}}^{n+1} + D_{i-\frac{1}{2}}^{n+1}\right), q_i = -\frac{\Delta t}{\Delta x^2} D_{i+\frac{1}{2}}^{n+1} \tag{13-2}$$

$$r_i = (\theta_w)_i^n - \frac{\rho_i}{\rho_w}[(\theta_i)_i^{n+1} - (\theta_i)_i^n] - \frac{\Delta t}{\Delta x}\left(k_{i+\frac{1}{2}}^{n+1} - k_{i-\frac{1}{2}}^{n+1}\right) \tag{13-3}$$

The matrix form of Eq. (13-1) is expressed as:

$$\boldsymbol{J}_\theta : \boldsymbol{\theta}_w = \boldsymbol{F}_\theta \tag{14}$$

where $\boldsymbol{J}_\theta$ is the Jacobian matrix for unfrozen water, $\boldsymbol{\theta}_w$ is the unfrozen water matrix, and $\boldsymbol{F}_\theta$ is the residual matrix for unfrozen water. Details of Eq. (14) are also shown in Appendix A.

The central difference utilizes internal nodes for computation, while values at the boundary nodes are provided using Dirichlet and Neumann boundary conditions, e.g.,

$$T|_\Gamma = T(x,t), \text{for Dirichlet boundary} \tag{15-1}$$

$$-\lambda \frac{\partial T}{\partial x}|_\Gamma = q_T(x,t), \text{for Neumann boundary} \tag{15-2}$$

where $\Gamma$ denotes the boundary and $q_T$ denotes the heat flux. Especially, the boundary conditions for thermal insulation and undrained samples are:

$$(T)_N = (T)_{N-1} \tag{16-1}$$

$$(\theta_w)_N = (\theta_w)_{N-1} \tag{16-2}$$

## 3.4 Numerical implementation of data-driven model

An effective algorithm is proposed to implement the ANN-model numerically, as shown in Fig. 4. The key steps are described as follows:

1. **Initialization**: At the beginning of each time step, the temperature and unfrozen water matrix are initially assumed to be the same as their values from the previous time step.

2. **Zone Classification**: The frozen and unfrozen zones are distinguished according to the freezing temperature $T_f$. The unfrozen water content at frozen nodes is then updated using the ANN model.

3. **Solution of the Hydraulic Equation**: Eq. (14) in the unfrozen zone is solved using the Newton-Raphson method. Since no ice formation occurs in the unfrozen region, the ice-related terms in Equations (13) and (14) are omitted. After solving the unfrozen water content, ice content is subsequently determined from the changes of liquid water content.

4. **Solution of the Thermal Equation**: Finally, Eq. (12) is solved using the Newton-Raphson method to obtain temperature distribution. During this process, steps 2 to 3 are repeated until both the temperature and unfrozen water content satisfy the required tolerance simultaneously, and the ANN model is recalled in each Newton iteration as described in step 2.

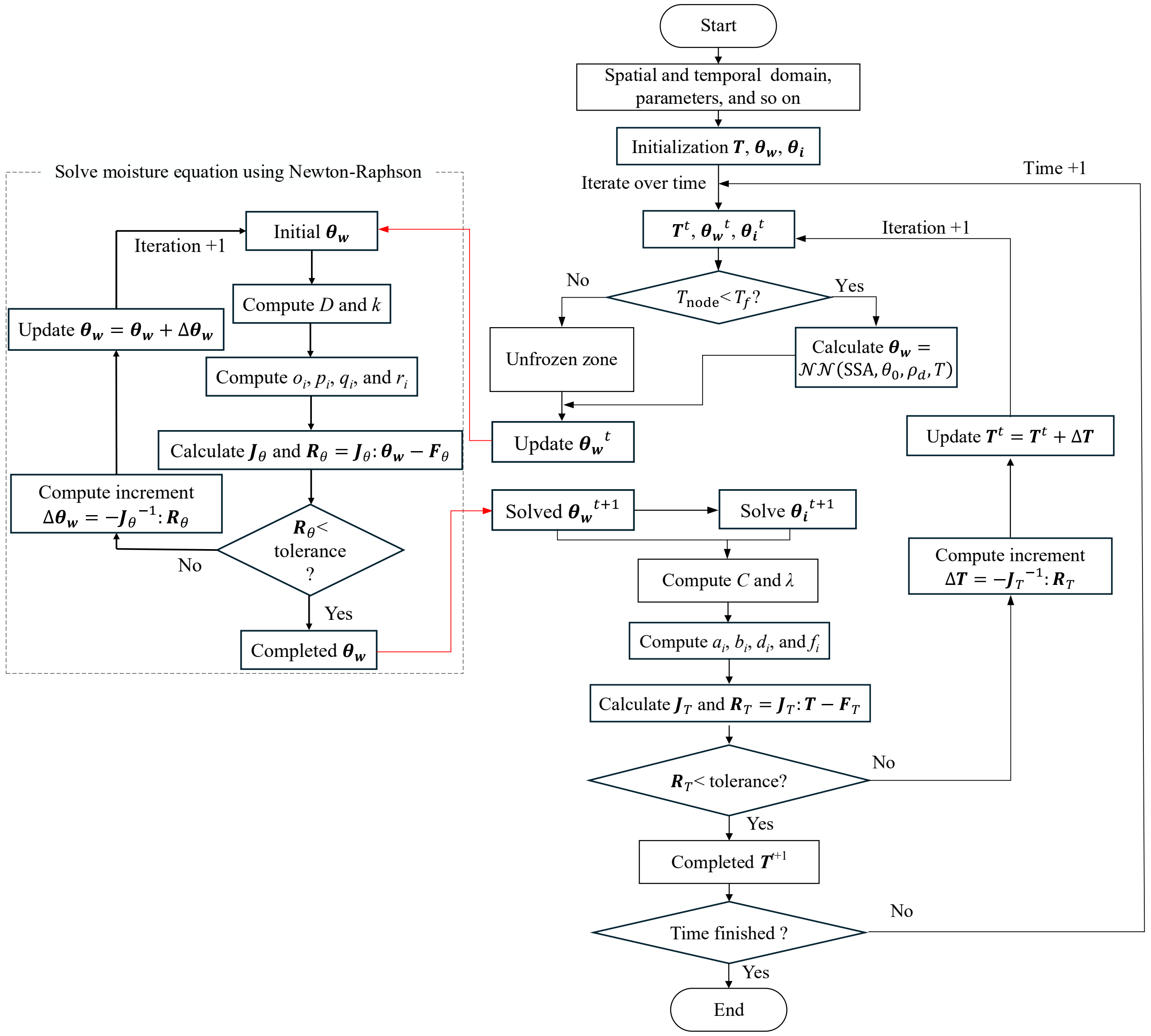

**Fig. 4**. Numerical implementation of the data-driven model for estimating unfrozen water content.

# 4. Model validation and results analysis

## 4.1 Experimental cases

Unidirectional freezing tests of soil columns conducted by Jame and Norum (1980) and Zhou et al. (2014) were selected to validate the proposed numerical algorithm. Jame and Norum (1980) utilized fine-grained silica flour samples with a dry density of 1330 kg/$m^3$ and a height of 30 cm. Two cases (case 1 and case 2) were modelled using the proposed method, as shown in Fig. 5(a). The horizontal soil columns were frozen from the left. The temperatures at the cold side were maintained at -5.9 °C and -5.2 °C for case 1 and case 2, respectively, while they were 4.25 °C and 5.0 °C, respectively, at the warm side. Both soil columns were insulated, and no external water was supplied. The initial mass water contents for case 1 and case 2 were 15% and 10.08%, respectively, and it was assumed to be homogeneous throughout the sample. Temperature and total water content during the testing were measured and reported along the sample length - see Jame and Norum (1980).

Zhou et al. (2014) conducted more detailed unidirectional freezing tests and reported the data of temperature, total water content, unfrozen water content, and ice content. The height of these vertical soil columns was 23.6 cm. The soil was composed of 30% sand, 69% silt, and 1% clay, classified as silt loam, with a dry bulk density of 1500 kg/$m^3$ and an initial porosity of 0.467. Two cases (case 3 and case 4) with different water contents were selected for numerical modelling, as shown in Fig. 5(b). The soil columns were frozen from the top side. The temperatures at the cold side were -4 °C and -4.2 °C for case 3 and case 4, respectively, while they were 3.6 °C and 4.1 °C at the warm side. In addition, the initial temperature was 3 °C for both experiments. Both soil columns were under insulated conditions and had no external water supply. The initial water contents for case 3 and case 4 were 32.5% and 22.5%, respectively.

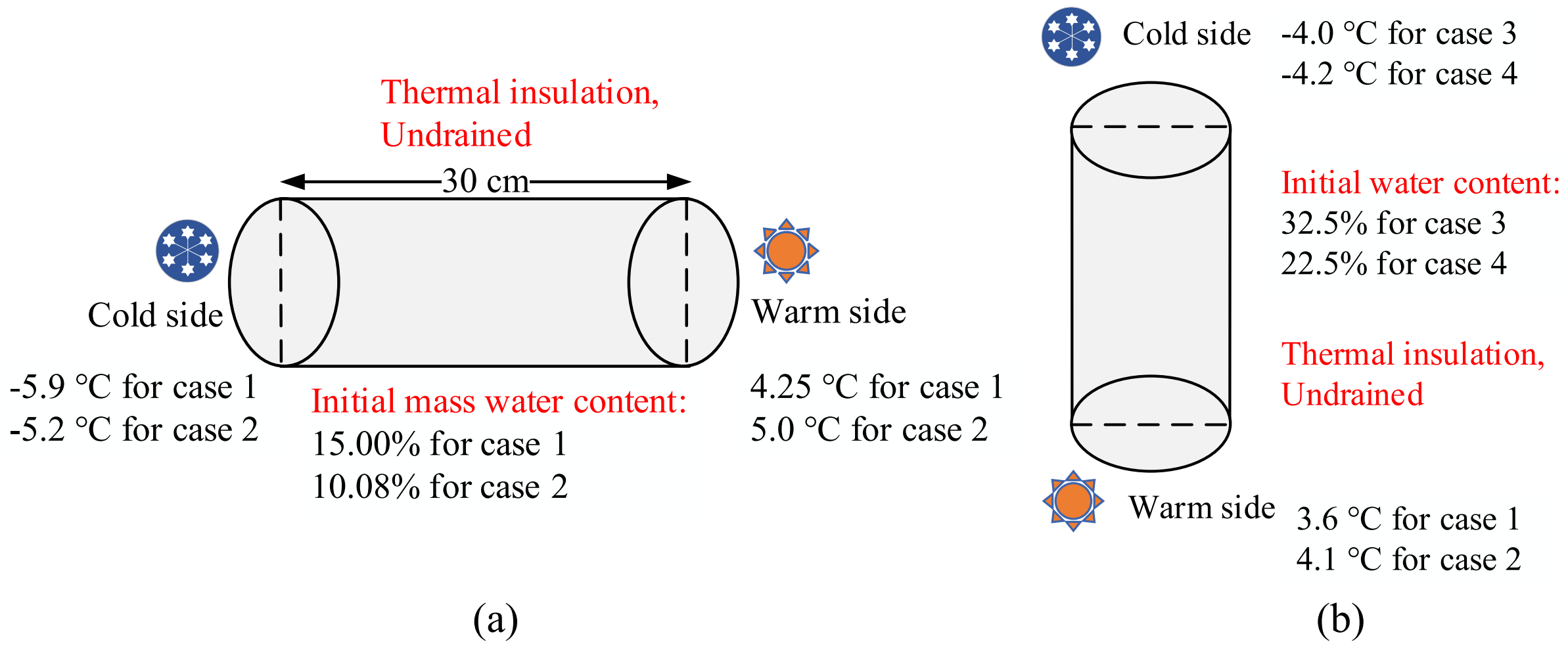

**Fig. 5**. Information of (a) case 1 and case 2 in Jame and Norum (1980), as well as (b) case 3 and case 4 in Zhou et al. (2014).

Table 2 summarizes model parameters used in numerical simulations. Among them, some parameters are constant values (i.e., $\rho_w$, $\rho_i$, $c_w$, $c_i$, $\lambda_w$, $\lambda_i$, and $L$) or can be found in the literature (i.e., $\rho_s$ and $\theta_s$). However, some parameters, such as $c_s$, $\lambda_s$, $\beta$, $k_s$, $\alpha$, and $m$, are not directly available, which were instead estimated from similar soil types or related references, e.g., Chen et al. (2022) and Tai et al. (2017). Besides, the freezing temperature is observed to be -0.22°C from the experiment of Zhou et al. (2014). Furthermore, to ensure the stability of spatial and temporal difference, the number of nodes $N$ should be relatively small while the number of time step $M$ should be large. By trial and error, we set $N$ less than 25. In addition, the time interval $\Delta t$ was configured as follows: 300 s in the first 4 hours, 1200 s in the next 20 hours, and finally 3600 s until the end of freezing, that is $M = 157$.

**Table 2**. Parameters used in numerical simulations.

| Parameter | Notation | Case 1 and case 2 (Jame and Norum, 1980) | Case 1 and case 2 (Zhou et al., 2014) | Unit |
|---|---|---|---|---|
| $\rho_s$ | Density of soil particles | 2600 | 2671 | kg/m$^3$ |
| $\rho_w$ | Density of water | 1000 | 1000 | kg/m$^3$ |
| $\rho_i$ | Density of ice | 917 | 917 | kg/m$^3$ |
| $c_s$ | Specific heat of soil particles | 1790 | 1790 | J/kg/K |
| $c_w$ | Specific heat of water | 4200 | 4200 | J/kg/K |
| $c_i$ | Specific heat of ice | 2100 | 2100 | J/kg/K |
| $\lambda_s$ | Thermal conductivity of soil particles | 1.18 | 1.18 | W/m/K |
| $\lambda_w$ | Thermal conductivity of water | 0.58 | 0.58 | W/m/K |
| $\lambda_i$ | Thermal conductivity of water | 2.31 | 2.31 | W/m/K |
| $L$ | Latent heat of water-ice | 334560 | 334560 | J/kg |
| $\beta$ | Thermal parameter | 0.9 | 0.3 | - |
| $\theta_r$ | Residual water content | 0.05 | 0.05 | - |
| $\theta_s$ | Saturated water content | 0.45 | 0.467 | - |
| $k_s$ | Saturated hydraulic conductivity | 1.2e-6 | 8.5e-8 | m/s |
| $\alpha$ | Parameter in VG model | 0.1 | 0.092 | 1/m |
| $m$ | Parameter in VG model | 0.75 | 0.70 | - |
| $T_f$ | Freezing temperature | -0.22 | -0.22 | °C |

## 4.2 Results and validation

### 4.2.1 Cases 1 and 2 – Horizontal freezing

In case 1 and case 2 the mass water content was translated into volumetric water content as follows:

$$m_t = \frac{\rho_w\theta_w+\rho_i\theta_i}{\rho_s(1-\theta_s)} \tag{17}$$

where $m_t$ is the mass of total water content.

Figs. 6(a) and (b) present the comparison between the predicted and the measured temperature and total mass water content for case 1 at freezing time of 12, 24, and 72 hours. The results demonstrate satisfactory agreements. As the freezing progresses, the frozen front moves to the right. Additionally, Fig. 6(c) presents the simulated unfrozen water and ice content. Within the frozen zone, the unfrozen water rapidly decreases to approximately the residual water content, which induces the water migration from unfrozen zone and consequently reduces the liquid water content therein. Simultaneously, the ice content in the frozen zone rises to exceed the initial water content.

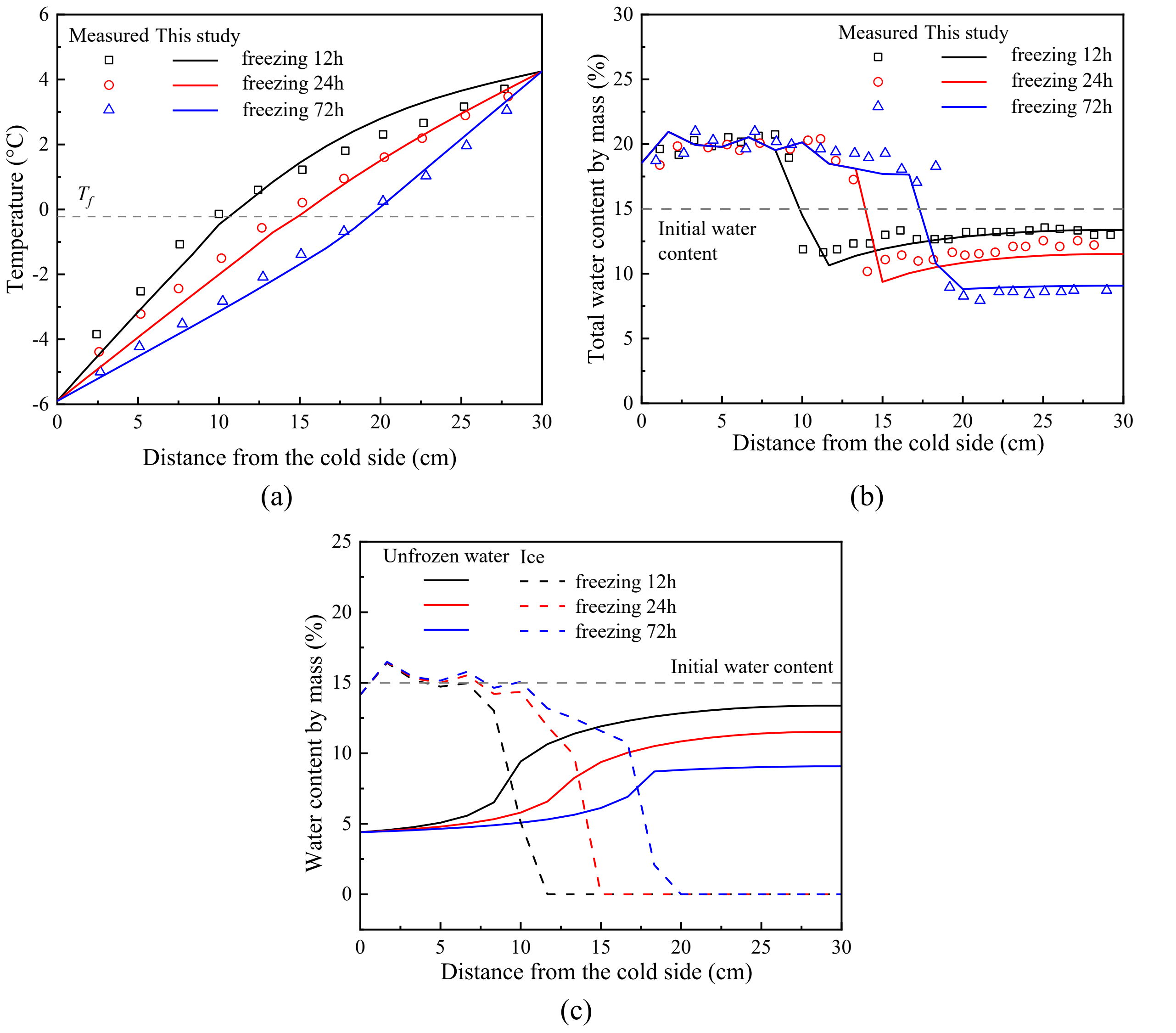

**Fig. 6**. Comparison results for case 1 (Jame and Norum, 1980) in (a) temperature; (b) total water content; (c) unfrozen water and ice content.

Fig. 7 compares the results of this study with the measured temperature and total mass water content for case 2, which also shows good agreements. Key features of the frozen soil, i.e., water accumulation at freezing front and water migration from warm side towards the frozen zone, are successfully reproduced by the simulation. However, a slight discrepancy between the measured and simulated total water content in the unfrozen zone is observed in Fig. 7(b) at the freezing time of 72 hours. This discrepancy may arise from the difference between the realistic and assumed hydraulic conductivities of soil. To this end, an investigation on the model parameters is conducted in Appendix B to understand their influence on simulation results. Besides, Fig. 7(b) depicts a more pronounced water accumulation, due to the increase in ice content (Fig. 7(c)), at the freezing front compared to case 1. For a larger initial water content with a higher seepage effect, liquid water is capable of transporting deeper into the frozen zone, resulting in a higher water content close to the cold side in Fig. 6(b). In contrast, a lower initial water content disrupts water seepage, thus the water tends to accumulate at the freezing front instead of migrating to the cold end.

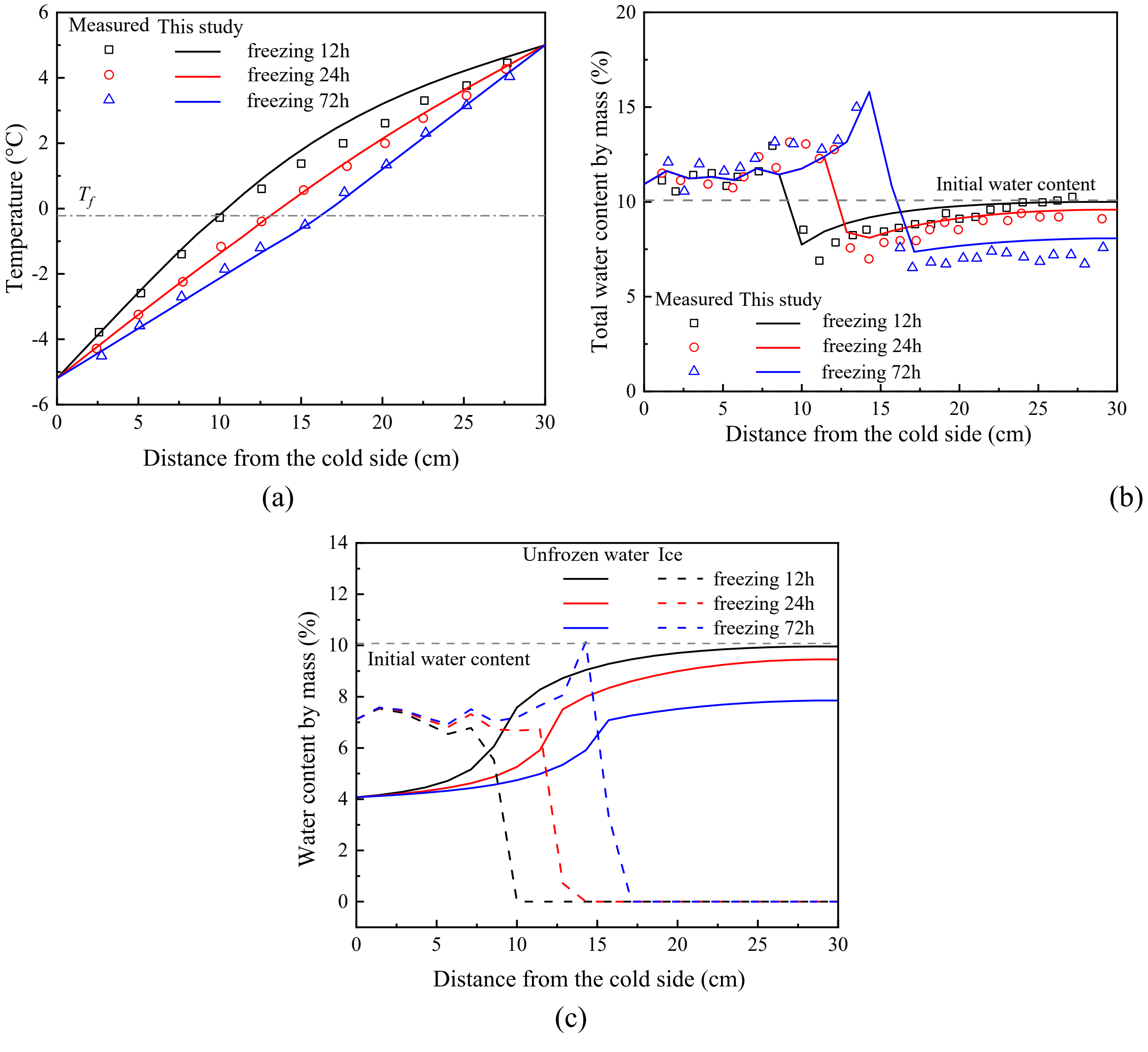

**Fig. 7**. Comparison results for case 2 (Jame and Norum, 1980) in (a) temperature; (b) total water content; (c) unfrozen water and ice content.

### 4.2.2 Cases 3 and 4 – Vertical freezing

Fig. 8 displays the comparison between the measured and simulated temperatures, total water content, unfrozen water content, and ice content for case 3 at freezing time of 1, 2, and 3 days, respectively.

The results agree generally well with the experimental data, although the calculated temperatures at a few points are slightly lower than the measured values. The reason for this discrepancy may be attributed to the inconsistencies between the thermal conductivity or hydraulic conductivity used in the simulations and those in the actual experiments. The error in the temperature calculations then lead to some discrepancies in the unfrozen water content, which subsequently cause slight errors in ice content and total water content.

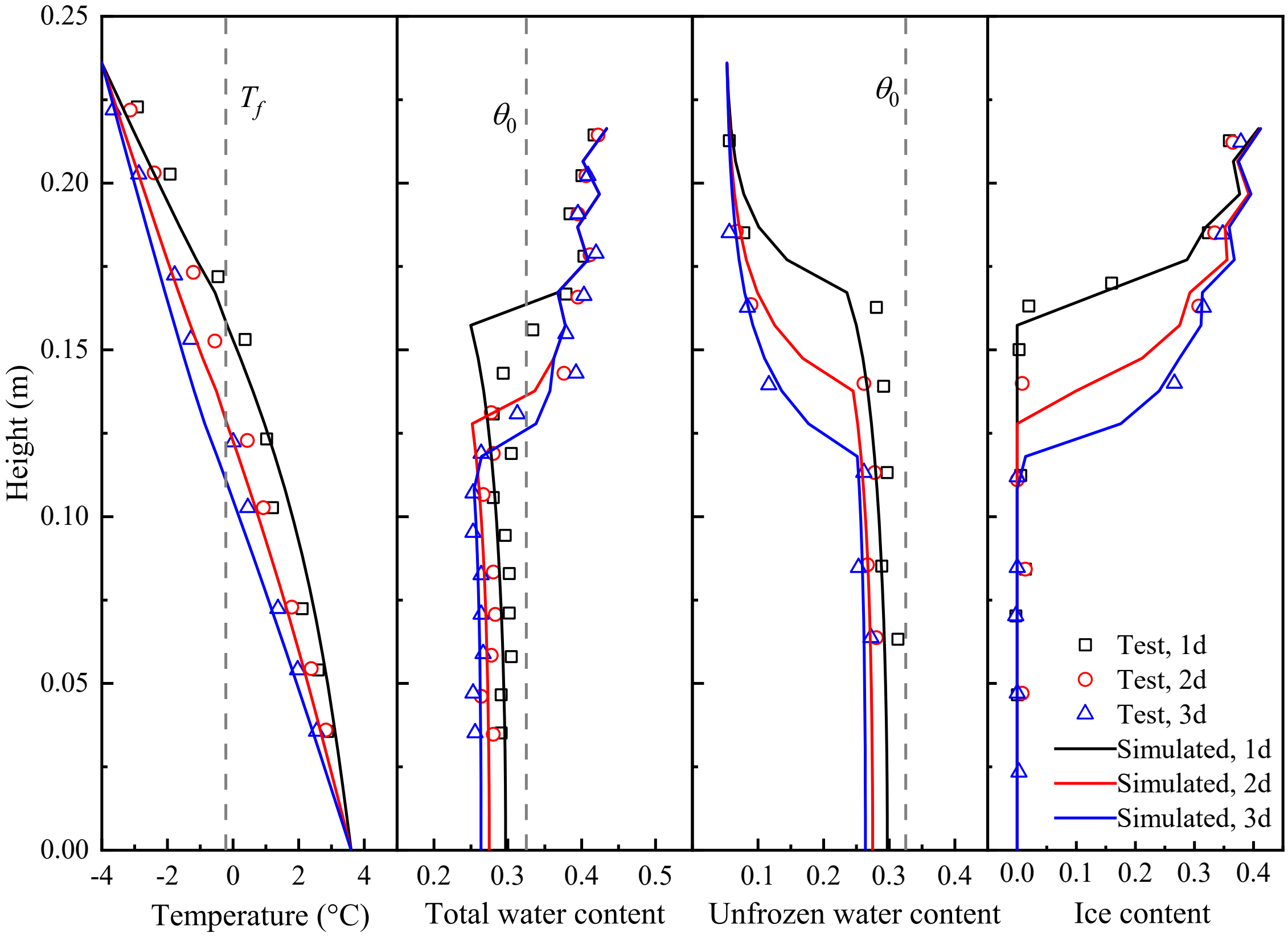

**Fig. 8**. Comparison results for case 3 (Zhou et al., 2014) in temperature, total water content, unfrozen water content, and ice content.

Fig. 9 compares the measured and simulated results for case 4. The numerical modelling also basically agrees well with the measured data. The simulated temperature distribution in Fig. 9 presents a closer alignment to the measured values, resulting in more accurate predictions of the unfrozen water content by the ANN model. Furthermore, a distinct peak for the total water content is observed, reaching a value of 0.386 at the freezing front. This peak is attributed to water migration from the unfrozen zone into the freezing front, where it subsequently freezes as illustrated in the ice content plot.

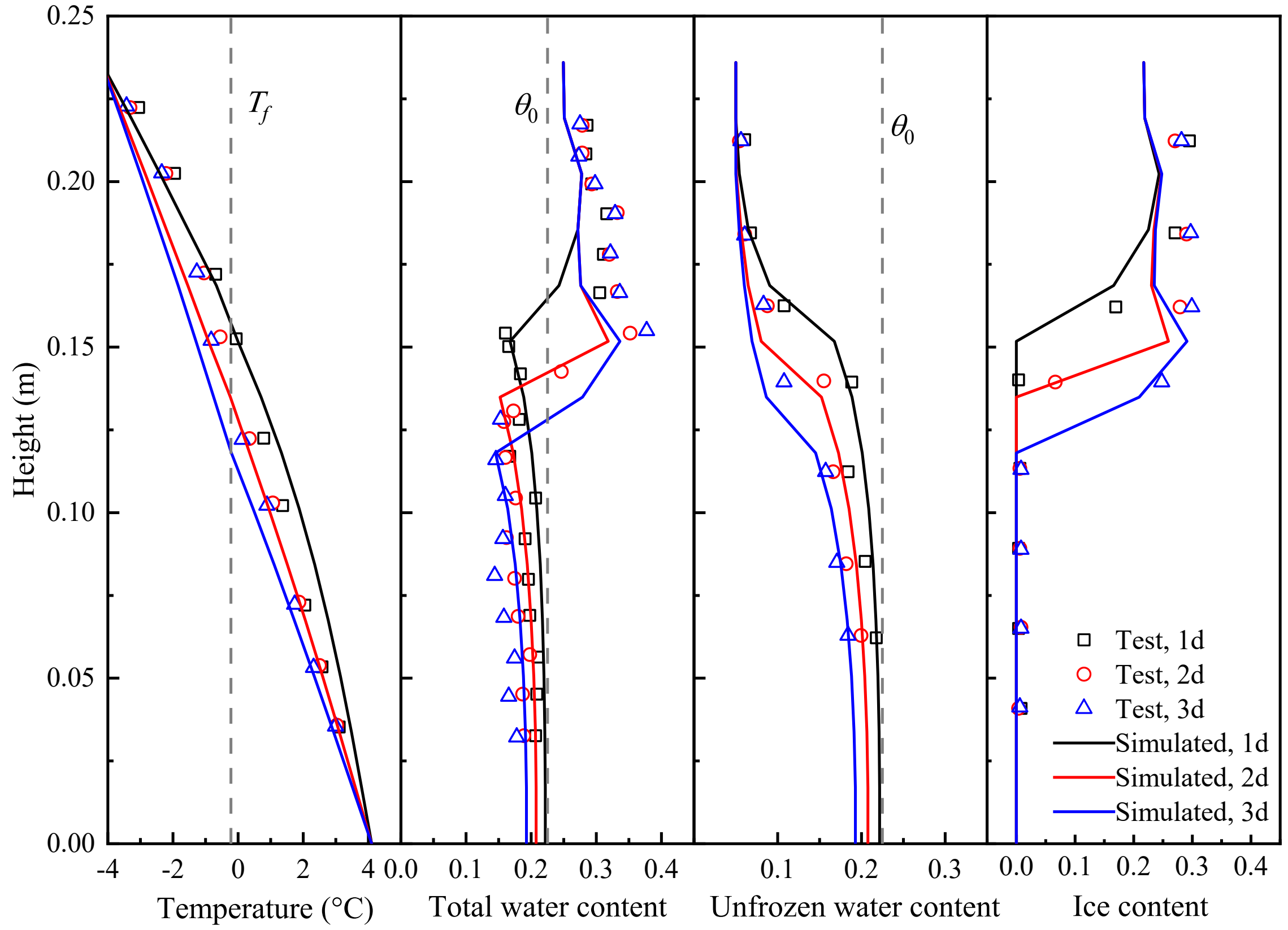

**Fig. 9**. Comparison results for case 4 (Zhou et al., 2014) in temperature, total water content, unfrozen water content, and ice content.

Fig. 10 presents the measured and simulated evolution of temperature, total water, unfrozen water, and ice content with freezing time at various heights for case 3. While the simulated temperature evolution in Fig. 10(a) generally follows the experimental trend, moderate discrepancies are observed, particularly at the initial stage. The simulated evolution of total water, unfrozen water, and ice content shows good agreements with the experimental data. It can be observed that when the freezing front arrives, the liquid water content rapidly decreases to the residual water content while simultaneously undergoing phase change to ice. At the same time, liquid water in the unfrozen zone decreases due to the seepage effect. Under such conditions, the total water content initially decreases and then rapidly increases to exceed the initial water content (including 9% volumetric expansion). As the freezing front passes away, the total water content remains almost constant with time, indicating that no water seepage into the frozen soil due to the resistance of ice.

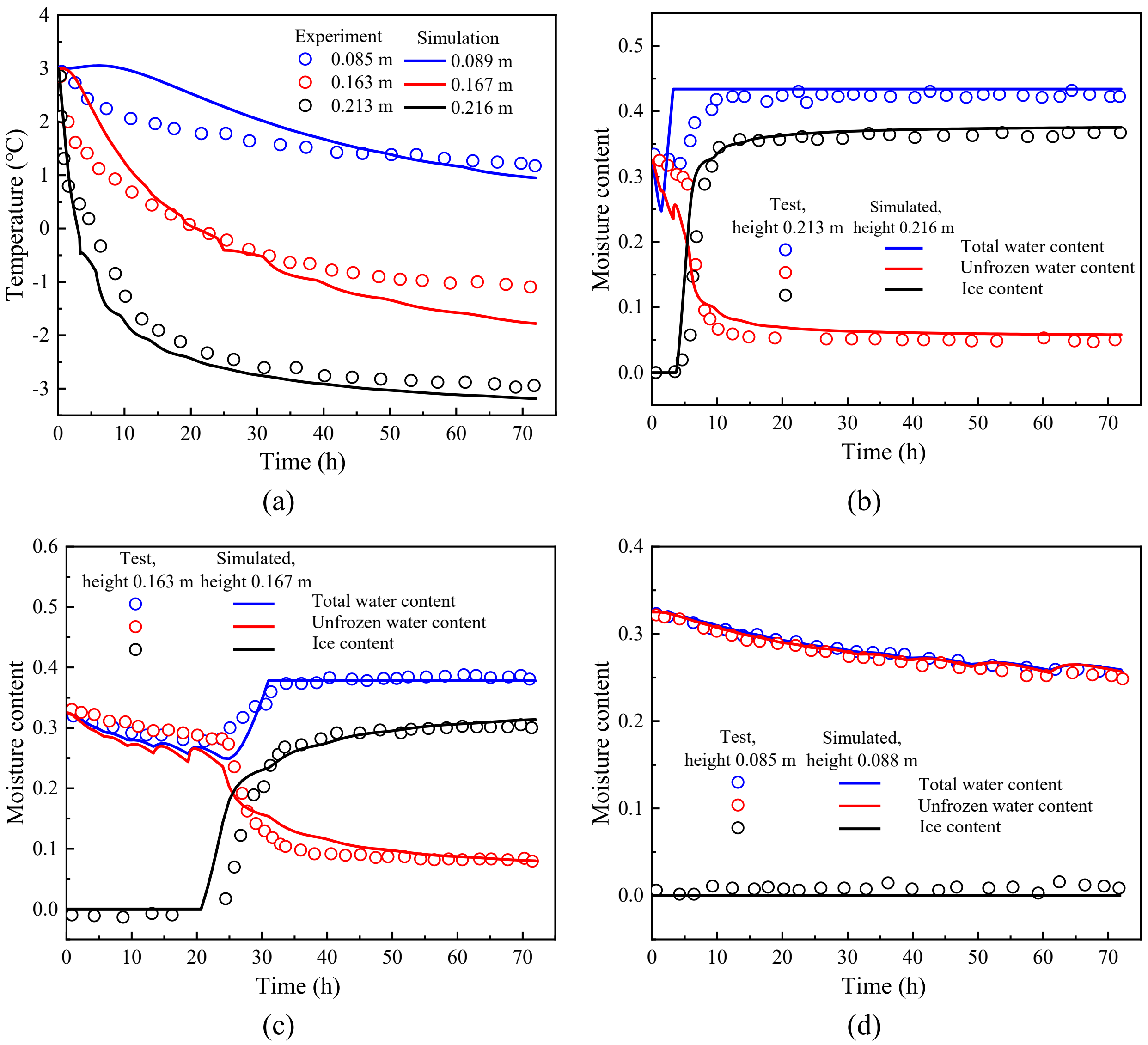

**Fig. 10**. Measured and simulated results with freezing time for case 3: (a) temperature at different heights, as well as total water, unfrozen water, and ice content at (b) height 0.213 m, (c) height 0.163 m, and (d) height 0.085 m.

Both experimental observations and simulation results reveal two processes during ice formation: (1) phase change from the original liquid water within the soil, and (2) accumulation and subsequent freezing of the migrated water from unfrozen zone. The phase change caused by the original liquid water can be computed by the ANN model. Additionally, the migrated water accumulates at the freezing front, where the accumulated water eventually freezes into ice. This explains why, at specific heights, the increase in ice content exceeds the corresponding decrease in unfrozen water content, as observed in Fig. 10. From a numerical simulation perspective, this process is controlled by the reduction of hydraulic conductivity. Konrad and Morgenstern (1980) and Zhou et al. (2014) pointed out the existence of a critical hydraulic conductivity threshold that prevents further water migration into frozen soil. In the frozen zone, hydraulic conductivity decreases significantly as ice forms, as described by Eq. (9-3). Consequently, migrated liquid water freezes at the bottom nodes of the frozen zone, aligning with the observed phenomenon of water accumulation at the freezing front and no water seeps into frozen zone.

## 4.3 Comparison with soil-freezing equations

Previous thermo-hydraulic coupling models for frozen soil typically rely on empirical or physical equations (Chen et al., 2022; Wang et al., 2024). To evaluate the effectiveness of the ANN approach, this section compares its performance with that of representative conventional equations. Hu et al. (2020) conducted a comprehensive review of existing methods for determining unfrozen water content in frozen soils and identified several equations with superior accuracy. Based on their findings, three of these equations were selected for comparison, as listed in Table 3. It should be noted that the parameters in Table 3 have been carefully fine-tuned to ensure optimal simulation performance across all four test cases.

**Table 3**. Three SFCCs for comparison.

| Expression | | Parameter | Reference |
|---|---|---|---|
| $\theta_w = \theta_r + (\theta_0 - \theta_r)\exp\left[-a\left(T - T_f\right)\right], T \le T_f$ | Eq. (17) | $a$ = -1.1 | Michalowski (1993) |
| $\theta_w = \theta_r - (\theta_0 - \theta_r)\frac{a}{T - a}, T \le T_f$ | Eq. (18) | $a$ = 0.5 | Westermann et al. (2011) |
| $\theta_w = a\|T\|^b, T \le T_f$ | Eq. (19) | $a$ = 0.07, $b$ = -0.47 | Anderson and Tice (1972) |

To compare the performance between various methods, we first define the relative error using Fréchet distance, as shown in Fig. 11(a). Each experimental point has a minimum distance to the simulation curve (i.e., Fréchet distance). The relative error is defined as the average Fréchet distance among all experimental points. Additionally, all variables are rescaled to the range of 0 ~ 1 to eliminate the influence of variable units. Fig. 11(b) compares the error predicted by the ANN model and the empirical equations among all cases. It can be observed that the ANN model outperforms other empirical equations with the lowest error in both temperature, liquid water, and ice content. Among three empirical equations, Michalowski (1993) and Westermann et al. (2011) present relatively large error, especially on the total water content. The power function form by Anderson and Tice (1972) performs slightly worse than the ANN model but is better than the other two empirical equations.

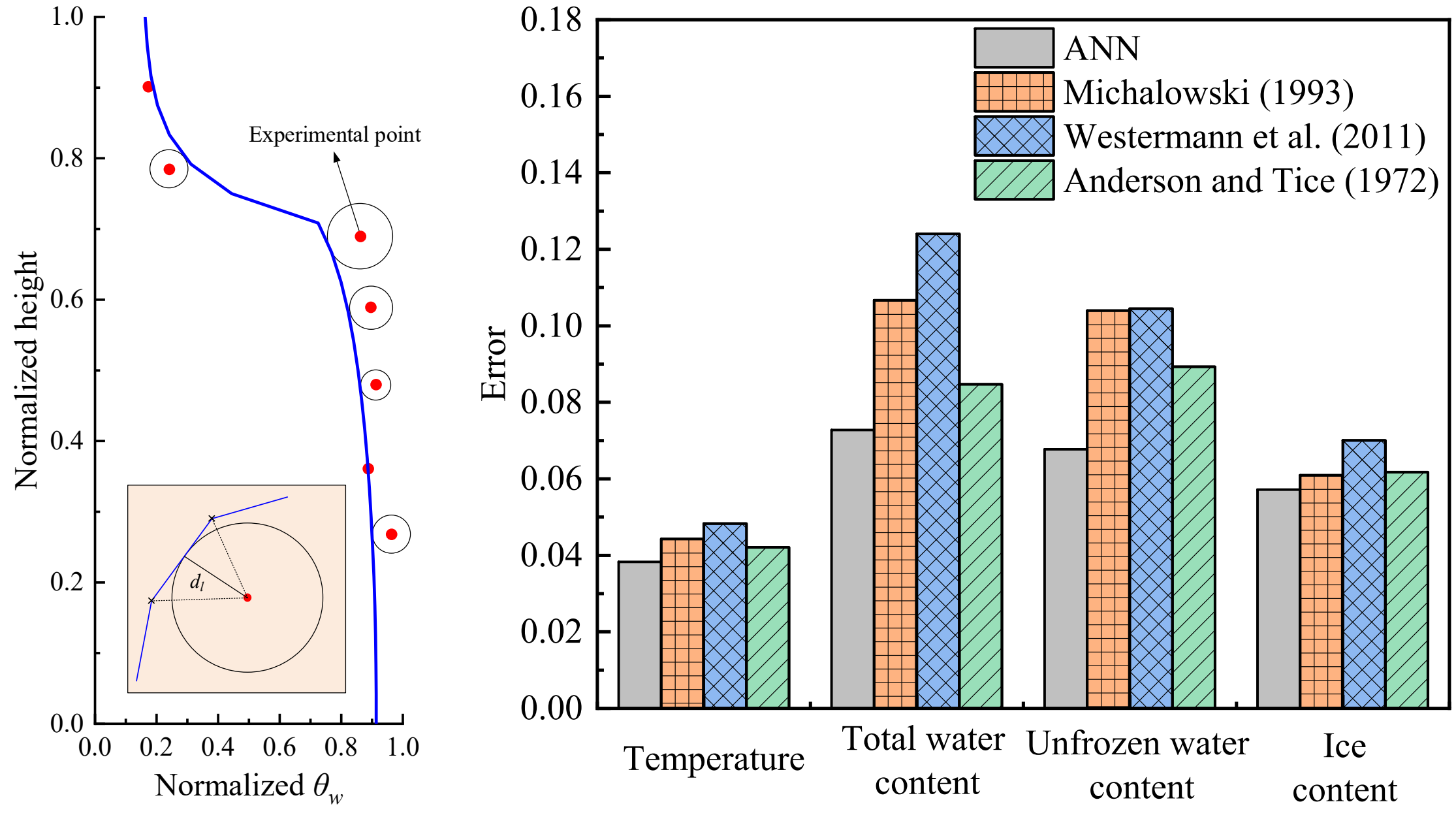

(a) (b)

**Fig. 11**. (a) Definition of Fréchet distance ($d_l$) between experimental points and model predictions; (b) Comparison of relative error between the ANN model with various equations.

Further, the comparison of the SFCC in cases 3 and 4 between the ANN model and empirical equations are presented in Fig. 12. The results demonstrate that the ANN model achieves the highest accuracy, closely matching the measured data. In contrast, the empirical equations proposed by Michalowski (1993) and Westermann et al. (2011) show relatively poor performance, failing to capture the evolution of unfrozen water content with temperature. The power function by Anderson and Tice (1972) performs better than the other two empirical equations but is not capable of accounting for the influence of initial water content. The advantages of the data-driven model include: 1) directly trained using the experimental data without any mathematical assumption; and 2) more universal because it is developed initially with multiple data and therefore does not require so intensive fine-tuning for specific ground and testing conditions.

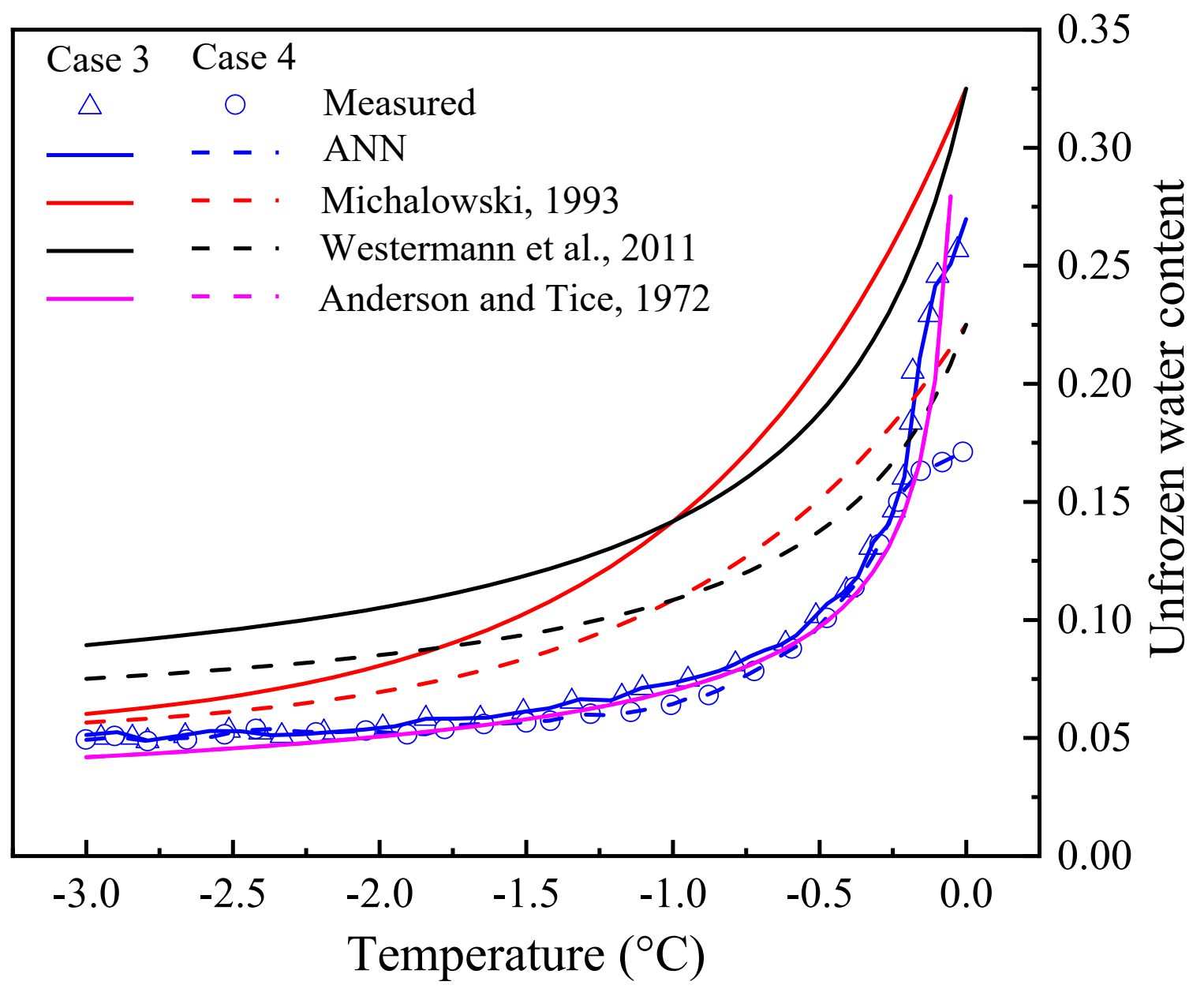

**Fig. 12**. Comparison of soil-freezing test between the ANN model with various equations for cases 3 and 4.

## 5. Conclusion

This study integrates a data-driven ANN model for estimating unfrozen water content into the thermo-hydraulic coupling simulation of frozen soils. The proposed numerical algorithm, validated through four experimental cases, demonstrates superior accuracy compared to empirical equations and other data-driven models (Ren et al., 2023) in modelling temperature evolution, liquid water migration, and ice formation. The ANN model, trained on 1,573 samples, takes specific surface area, dry density, initial water content, and temperature as inputs to predict unfrozen water content. With $R^2$ values of 98.6% and 95.9% for training and testing sets, respectively, the model effectively captures the evolution of unfrozen water content under subfreezing temperatures.

Then, a numerical algorithm was proposed to implement the data-driven model into the heat-moisture coupling simulation of frozen soil, which follows three key steps: (1) identifying frozen and unfrozen zones based on freezing temperature, (2) updating unfrozen water content in frozen nodes using the ANN model, and (3) solving the discretized moisture and heat equations iteratively via Newton-Raphson method. The proposed method was validated by comparing with data of temperature, total water, unfrozen water, and ice content obtained from several freezing experiments.

This work implements the data-driven model for estimating the unfrozen water content into the heat-moisture coupling. Future work may focus on extending data-driven approaches to other key soil properties, such as thermal and hydraulic conductivity, for more comprehensive numerical simulations.

# Appendix A

For Eq. (12) $\boldsymbol{J}_T{:}\,\boldsymbol{T} = \boldsymbol{F}_T$, each item is shown as:

$$\boldsymbol{J}_T = \begin{bmatrix} a_2 & b_2 & d_2 & & & & \\ & a_3 & b_3 & d_3 & & & \\ & & \cdots & \cdots & \cdots & & \\ & & & a_i & b_i & d_i & \\ & & & & \cdots & \cdots & \cdots \\ & & & & a_{N-1} & b_{N-1} & d_{N-1} \end{bmatrix} \tag{A-1}$$

$$\boldsymbol{T} = [0 \quad (T)_2^{n+1} \quad (T)_3^{n+1} \quad \cdots \quad (T)_i^{n+1} \quad \cdots \quad (T)_{N-1}^{n+1} \quad 0]^{\text{Transpose}} \tag{A-2}$$

$$\boldsymbol{F}_T = [0 \quad f_2 \quad f_3 \quad \cdots \quad f_i \quad \cdots \quad f_{N-1} \quad 0]^{\text{Transpose}} \tag{A-3}$$

For Eq. (14) $\boldsymbol{J}_\theta{:}\,\boldsymbol{\theta_w} = \boldsymbol{F}_\theta$, each item is shown as:

$$\boldsymbol{J}_\theta = \begin{bmatrix} o_2 & p_2 & q_2 & & & & \\ & o_3 & p_3 & q_3 & & & \\ & & \cdots & \cdots & \cdots & & \\ & & & o_i & p_i & q_i & \\ & & & & \cdots & \cdots & \cdots \\ & & & & o_{N-1} & p_{N-1} & q_{N-1} \end{bmatrix} \tag{A-4}$$

$$\boldsymbol{\theta_w} = [0 \quad (\theta_w)_2^{n+1} \quad (\theta_w)_3^{n+1} \quad \cdots \quad (\theta_w)_i^{n+1} \quad \cdots \quad (\theta_w)_{N-1}^{n+1} \quad 0]^{\text{Transpose}} \tag{A-5}$$

$$\boldsymbol{F}_\theta = [0 \quad r_2 \quad r_3 \quad \cdots \quad r_i \quad \cdots \quad r_{N-1} \quad 0]^{\text{Transpose}} \tag{A-6}$$

# Appendix B

## B.1 Effect of thermal-related parameters

The sensitivity of thermal-related parameters, i.e., $\lambda_s$, $c_s$, and $\beta$, are investigated using case 2 as an example (other parameters set the default values in Table 2). As shown in Fig. B1(a) and (b), the temperature distribution exhibits minimal variation with substantial increase in $\lambda_s$ and $c_s$, indicating that the model is relatively insensitive to changes in these parameters. This reflects a limitation of the traditional weighted mean model for calculating thermal conductivity (i.e., when $\beta$ is omitted in Eq. (7)), as it lacks adaptability across different soil types. In contrast, Fig. B1(c) presents that the temperature rises obviously with the decreasing $\beta$ values. This highlights the effectiveness of the improved weighted mean model to calculate thermal conductivity (Eq. (7)), which enhances its adaptability to a broader range of soil conditions.

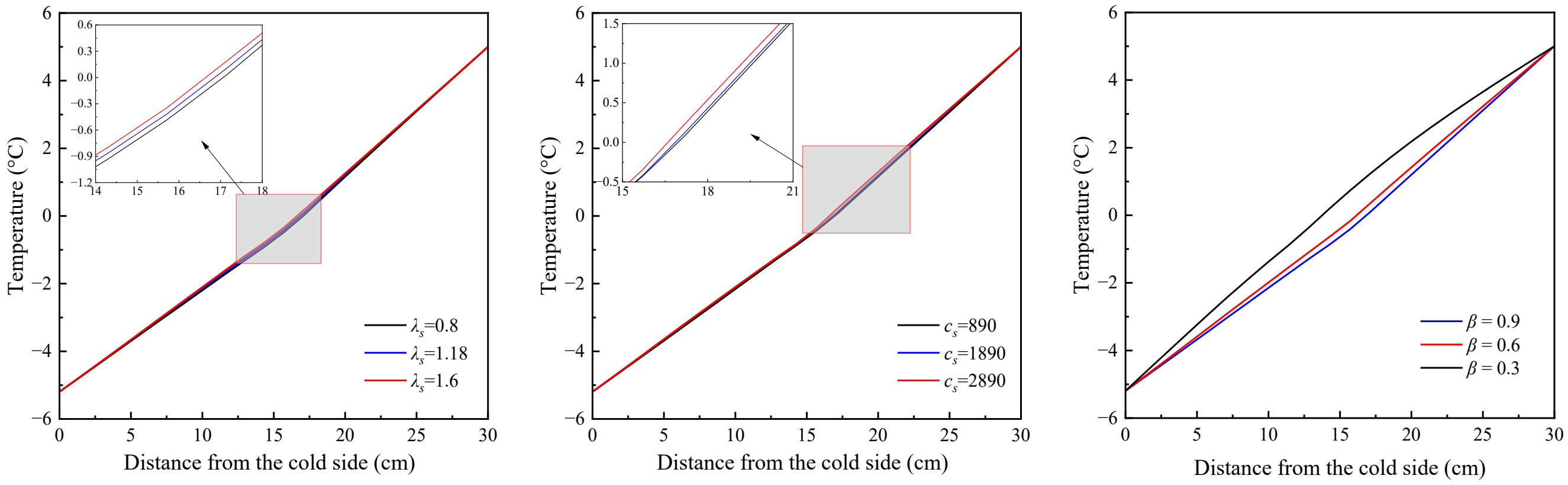

(a) (b) (c)

**Fig. B1**. Sensitivity of thermal-related parameters on temperature: (a) $\lambda_s$, (b) $c_s$, and (c) $\beta$.

## B.2 Effect of hydraulic-related parameters

The sensitivity of hydraulic-related parameters, i.e., $k_s$, $\alpha$, and $m$, are examined, as shown in Figs. B2 to B4. As depicted in Fig. B2, a higher value of $k_s$ indicates greater water permeability, allowing more liquid water from the unfrozen zone to migrate into the frozen zone, thereby increasing the ice content. When $k_s$ increases from 2.2e-7 to 1.2e-6, the water permeability improves, yet it remains insufficient for water to reach the top end of the frozen zone, which in turn enhances water accumulation at the freezing front. However, when $k_s$ further increases to 5.2e-6, the seepage capacity becomes strong enough to allow water to penetrate to the upper regions of the frozen zone, where it subsequently freezes into ice and shows a higher ice content in the top of frozen zone.

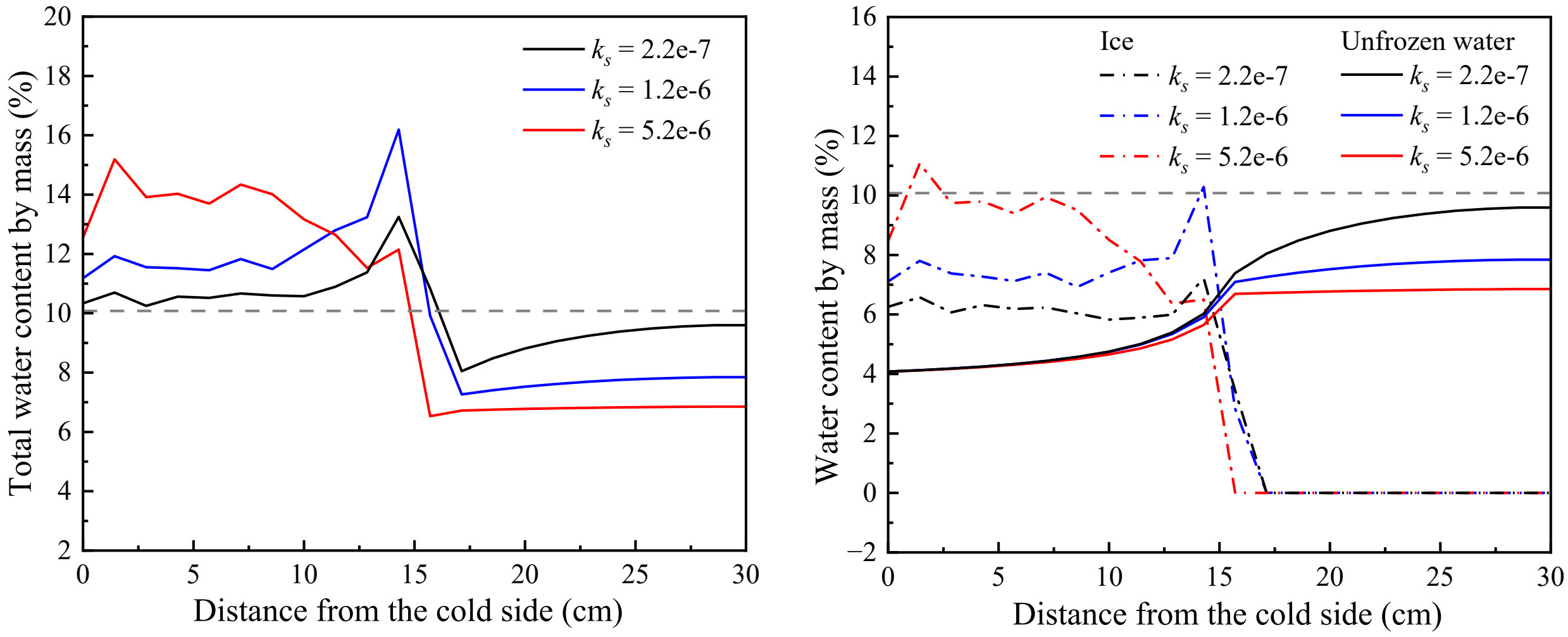

(a) (b)

**Fig. B2**. Sensitivity of $k_s$ on (a) total water content, and (b) unfrozen water and ice content.

Parameter $\alpha$ is related to the calculation of SWRC, thereby influencing the water distribution. As shown in Fig. B3, a reduction in $\alpha$ value enhances the water permeability, resulting in less liquid water in the unfrozen zone and higher ice content in the frozen zone. When $\alpha$ decreases from 0.3 to 0.1, the total water content presents a more pronounced water accumulation at the freezing front. However, with a further decrease to $\alpha = 0.035$, water permeability increases significantly, allowing liquid water to migrate into the upper portion of the frozen zone, rather than accumulating at the freezing front.

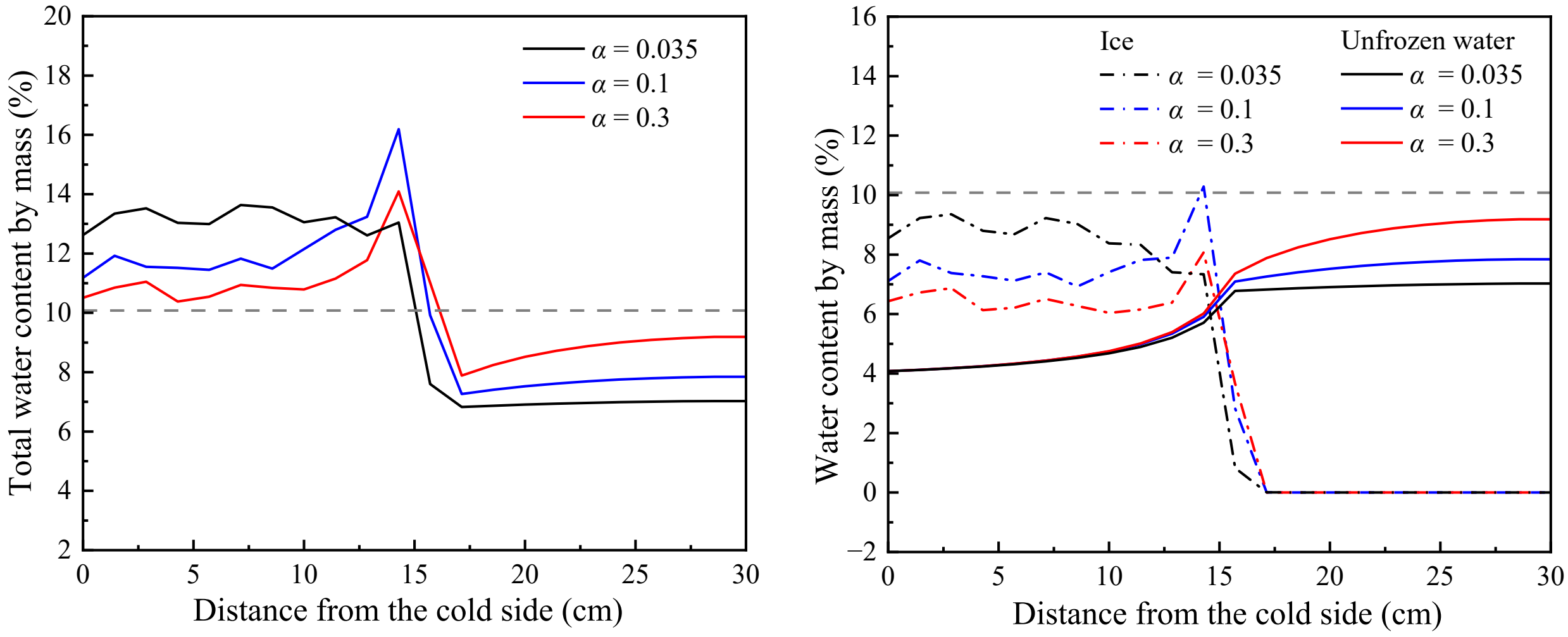

(a) (b)

**Fig. B3**. Sensitivity of $\alpha$ on (a) total water content, and (b) unfrozen water and ice content.

Similarly, the SWRC is also influenced by parameter $m$, in which way to affect the water distribution.

The impact of $m$ (ranging 0 to 1) is illustrated in Fig. B4. As $m$ increases from 0.45 to 0.75, water permeability improves, promoting greater water migration from the unfrozen to the frozen zone. However, when $m$ further increases to 0.96, water permeability decreases, leading to reduced water migration. These results indicate that extreme values of $m$ (approach 1 or 0) correspond to lower seepage capacity, while there exists an intermediate $m$ (depending on soil type) that maximizes water permeability. Additionally, the variation in $m$ alone is insufficient to facilitate liquid water seepage to the top of the frozen zone, and instead, water primarily accumulates at the freezing front.

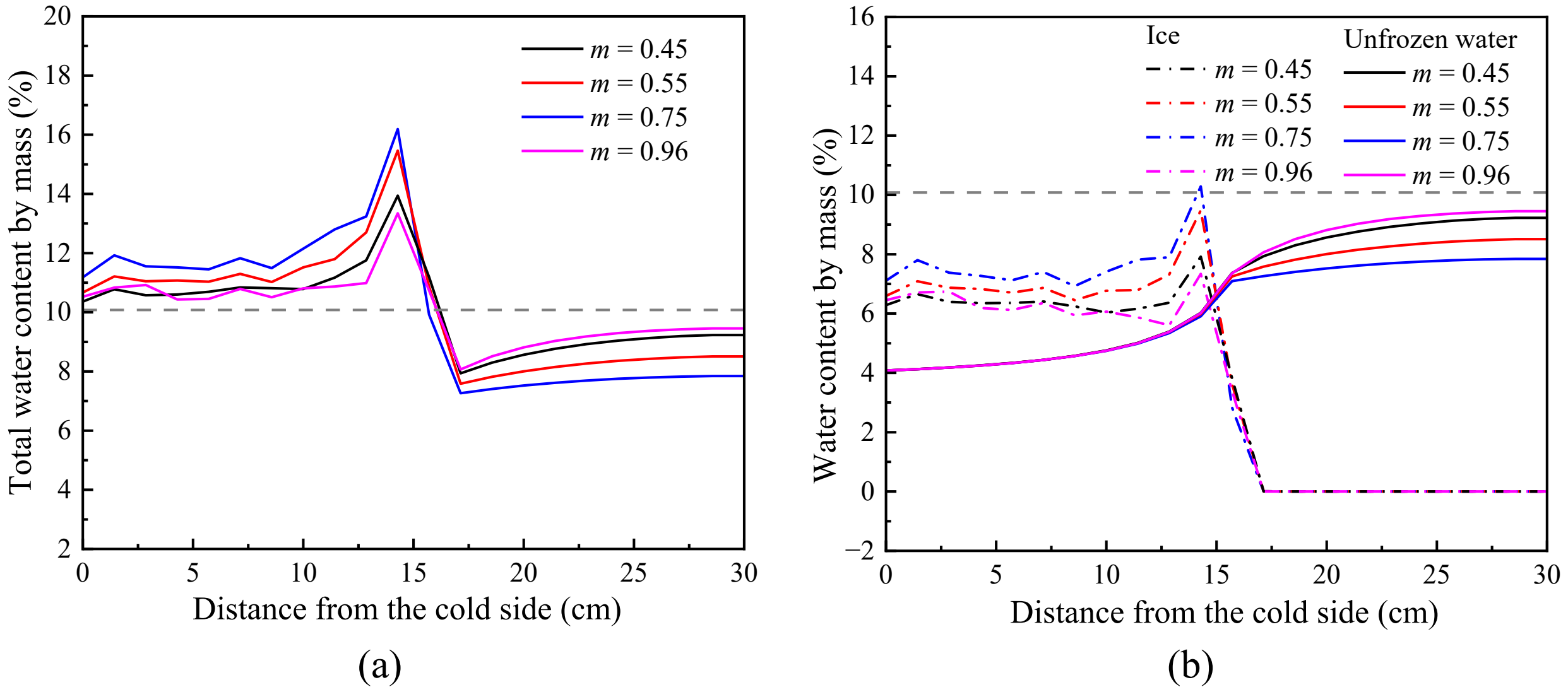

**Fig. B4**. Sensitivity of $m$ on (a) total water content, and (b) unfrozen water and ice content.

# Acknowledgement

The authors would like to acknowledge the funding support by the China Scholarship Council (CSC).